\documentclass[usenatbib]{mn2e}
\usepackage{psfig}
\usepackage{rotating}
\usepackage{multirow}
\usepackage{amssymb}
\usepackage{hyperref}
\usepackage{color}

\title[Near-Infrared and Weak-Lensing Scaling Relation] {LoCuSS: The Near-Infrared Luminosity and Weak-Lensing Mass Scaling Relation of Galaxy Clusters}

\author[Mulroy et al.]
{Sarah L.\ Mulroy$^{1,\star}$,
Graham P.\ Smith$^{1}$,
Chris P.\ Haines$^{1,2,3}$,
Daniel P.\ Marrone$^{2}$,
\and 
Nobuhiro Okabe$^{4}$,
Maria J.\ Pereira$^{2}$,
Eiichi Egami$^{2}$,
Arif Babul$^{5}$,
\and
Alexis Finoguenov$^{6,7}$,
and
Rossella Martino$^{8,9}$
\vspace{2mm}\\
	$^1$School of Physics and Astronomy, University of Birmingham, Edgbaston, Birmingham, B15 2TT, UK\\
	$^2$Steward Observatory, University of Arizona, 933 North Cherry Avenue, Tucson, AZ 85721, USA\\
	$^3$Departamento de Astronom\'{i}a, Universidad de Chile, Casilla 36-D, Correo Central, Santiago, Chile\\
    $^4$Kavli Institute for the Physics and Mathematics of the Universe (WPI), Todai Institutes for Advanced Study,\\ University of Tokyo, 5-1-5 Kashiwanoha, Kashiwa, Chiba 277-8583, Japan\\
	$^5$Department of Physics and Astronomy, University of Victoria, 3800 Finnerty Road, Victoria, BC V8P 1A1, Canada\\
	$^6$Department of Physics, University of Helsinki, Gustaf H\"{a}llstr\"{o}min katu 2a, FI-0014 Helsinki, Finland\\
	$^7$Center for Space Science Technology, University of Maryland, Baltimore County, 1000 Hilltop Circle, Baltimore, MD 21250, USA\\
    $^8$Dipartimento di Fisica, Universit\`a degli Studi di Roma ``Tor Vergata'', via della Ricerca Scientifica 1, 00133, Roma,  Italy\\
    $^9$Laboratoire AIM, IRFU/Service dÕAstrophysique -CEA - CNRS, Bt. 709, CEA-Saclay, 91191 Gif-sur-Yvette Cedex, France\\
        $^\star$Email: smulroy@star.sr.bham.ac.uk
}

\begin{document}

\date{Accepted, Received}

\pagerange{\pageref{firstpage}--\pageref{lastpage}} \pubyear{2013}

\maketitle

\label{firstpage}

\begin{abstract}
We present the first scaling relation between weak-lensing galaxy cluster mass, $M_{\rm WL}$, and near-infrared luminosity, $L_K$. Our results are based on 17 clusters observed with wide-field instruments on Subaru, the United Kingdom Infrared Telescope, the Mayall Telescope, and the MMT. We concentrate on the relation between projected 2D weak-lensing mass and spectroscopically confirmed luminosity within 1Mpc, modelled as $M_{\rm WL} \propto L_{K}^b$, obtaining a power law slope of $b=0.83^{+0.27}_{-0.24}$ and an intrinsic scatter of $\sigma_{lnM_{\rm WL}|L_{K}}=10^{+8}_{-5}\%$. Intrinsic scatter of $\sim10\%$ is a consistent feature of our results regardless of how we modify our approach to measuring the relationship between mass and light. For example, deprojecting the mass and measuring both quantities within $r_{500}$, that is itself obtained from the lensing analysis, yields $\sigma_{lnM_{\rm WL}|L_{K}}=10^{+7}_{-5}\%$ and $b=0.97^{+0.17}_{-0.17}$.  We also find that selecting members based on their $(J-K)$ colours instead of spectroscopic redshifts neither increases the scatter nor modifies the slope.  Overall our results indicate that near-infrared luminosity measured on scales comparable with $r_{500}$ (typically 1Mpc for our sample) is a low scatter and relatively inexpensive proxy for weak-lensing mass. Near-infrared luminosity may therefore be a useful mass proxy for cluster cosmology experiments.
\end{abstract}


\section{Introduction}

The growth rate and internal structure of galaxy clusters are sensitive to the cosmological model. Clusters are therefore well established cosmological tools that hold much promise for ongoing and imminent cosmological studies, including those that aim to measure the dark energy equation of state \citep[and references therein]{2011ARA&A..49..409A}. Clusters are tracers of the high mass end of the mass function and so to test cosmological models against observations requires an accurate measurement of the cluster halo mass. As the mass of clusters is dominated by dark matter, this quantity cannot be measured directly and generally requires extensive observations and modelling.

The importance and complexity of cluster mass measurements are among the key motivations for studying scaling relations between mass $M$ and another observable $O$, or ``mass proxy''. The form of these relations is motivated by predictions from self-similarity \citep{1986MNRAS.222..323K} that they are power laws, parameterised by normalisation $a$, slope $b$, and intrinsic scatter $\sigma_{\ln M|O}$. An ideal scaling relation has low intrinsic scatter, while an ideal observable is inexpensive to measure and preferably obtainable from shallow survey data. Also important are a clear understanding of the relationship between the measured mass and the ``true'' mass, and minimal covariance between $M$ and $O$.

Most scaling relation studies are based on X-ray observations, and thus assume that the intracluster medium is in hydrostatic equilibrium with the cluster potential. Gas mass, $M_{\rm gas}$, and X-ray temperature, $T_X$, have been shown to be related to the hydrostatic mass of clusters with intrinsic scatter of $\sim10\%$ and $\sim15-20\%$ respectively \citep[e.g.][]{2007A&A...474L..37A,2010MNRAS.406.1773M}. The pseudo-pressure of the intracluster gas, namely $Y_X=T_X\,.\,M_{\rm gas}$, was predicted by simulations to be related to hydrostatic mass with an intrinsic scatter as low as $5\%$ \citep{2006ApJ...650..128K}, however observations suggest a figure closer to $\gtrsim 15 \%$ 
\citep[e.g.][]{2007A&A...474L..37A,2010MNRAS.406.1773M,2014arXiv1406.6831M}.

Following several early exploratory studies \citep{Smail1997,Hjorth1998,Smith2005,2007A&A...470..449B}, scaling relation studies based on gravitational lensing mass measurements have developed rapidly in the last few years. The advantage of lensing mass measurement is that it makes no assumption about the dynamical and hydrostatic state of the cluster, although it has irreducible scatter of $\sim20-30\%$ due to projection effects and uncorrelated large-scale structure along the line of sight \citep[e.g.][]{2010A&A...514A..93M,2011ApJ...740...25B,2012MNRAS.421.1073B,2012NJPh...14e5018R}. Lensing-based results generally agree with X-ray-based studies that $M_{gas}$ is the lowest scatter X-ray mass proxy, with $\sim10-15\%$ intrinsic scatter \citep[e.g.][]{2010ApJ...721..875O,2013ApJ...767..116M}, with $Y_{X}$ presenting $\sim20-25\%$ scatter \citep[e.g.][]{2010ApJ...721..875O,2013ApJ...767..116M}. Recent measurements of the scaling relation between weak-lensing mass and the integrated Compton parameter, $Y_{SZ}$, find intrinsic scatter of $\sim10-20\%$ \citep{2012ApJ...754..119M,Hoekstra2012}, in broad agreement with Sunyaev-Zeldovich effect studies that employ hydrostatic mass estimates \citep[e.g.][]{2008ApJ...675..106B,2011ApJ...738...48A}.

The integrated optical/near-infrared luminosity of the cluster galaxies can also be used as a mass proxy. $K$-band luminosity is a well-known and reliable tracer of the stellar mass in galaxies, as it is sensitive to old stars and relatively insensitive to more recent star formation and dust extinction \citep{1998MNRAS.297L..23K}. Several studies have investigated near-infrared luminosity, finding that the $M-L_{K}$ scaling relation has a scatter of $\gtrsim30\%$ \citep[e.g.][]{2003ApJ...591..749L,2004ApJ...610..745L,2004AJ....128.2022R,2004AJ....128.1078R,2007ApJ...663..150M}. They have all used either dynamical or X-ray mass measurements. In contrast, strong- and weak-lensing studies of clusters report that near-infrared luminosity traces the density and structure of clusters to good accuracy \citep{2003ApJ...598..804K,Smith2005,Richard2010}. These results suggest that the relationship between weak-lensing mass and near-infrared luminosity may have a lower scatter than that between X-ray/dynamical mass and near-infrared luminosity.

In this article we present a pilot study of the scaling relation between weak-lensing mass and $K$-band luminosity for a sample of 17 clusters at $0.15\le z \le 0.3$. We summarise the gravitational weak-lensing masses and calculate the $K$-band luminosities in \S\ref{sec:data}. The results are presented in \S\ref{sec:results}, compared with other published results in \S\ref{sec:discussion}, and our findings summarised in \S\ref{sec:summary}. All photometric measurements are relative to Vega, and we assume $\Omega_{\rm M}=0.3$, $\Omega_\Lambda=0.7$ and $H_0=70\,{\rm km\,s^{-1}\,Mpc^{-1}}$. In this cosmology, at the average cluster redshift, $\langle z \rangle=0.23$, 1 arcsec corresponds to a physical scale of 3.67 kpc.

\section{Data and Analysis}\label{sec:data}

\subsection{Sample}\label{sec:sample}

We study a sample of 17 X-ray luminous clusters at $0.15<z<0.3$ (Table~\ref{tab:sample}) that have featured in a series of papers from the Local Cluster Substructure Survey (LoCuSS\footnote{\url{http://www.sr.bham.ac.uk/locuss}}). They are those with weak-lensing masses published in \citet[][see Table 6]{2010PASJ...62..811O} for which we have near-infrared observations of the cluster galaxies \citep{2009ApJ...704..126H}. As such, they were selected without reference to their X-ray morphology and temperature structure, and yielded a satisfactory weak-shear profile fit to a \citet{1997ApJ...490..493N} density profile. We will consider whether restricting to this sub-sample introduces any bias into our results in a future paper that will consider the full ``High-$L_X$'' LoCuSS sample.

\begin{table*}
\caption{Sample}\label{tab:sample}
	\begin{center}
		\tabcolsep=0.8mm
			\begin{tabular}{ l c c c c c c c c c }
		      \hline
		      \hline
			Name & Redshift & RA & Dec & $\rm N_{gal}$ & Completeness & $M_{2D}(< \rm 1Mpc)$ & $L_{K}(< \rm 1Mpc)$ & $M_{500}$ & $L_{K}(<r_{500})$\\
			& & [J2000] & [J2000] & ($<1 \rm Mpc$) & (\% $<1 \rm Mpc$) & $(10^{14}M_{\odot})$ & $(10^{12}L_{\odot})$ & $(10^{14}M_{\odot})$ & $(10^{12}L_{\odot})$\\
		      \hline
		      ABELL0068 & 0.2546 & 00 37 05.28 & $+$09 09 10.8 & 49 & 58 & $7.66^{+2.17}_{-2.17}$ & $13.45^{+2.19}_{-2.19}$ & $4.17^{+1.23}_{-1.07}$ & $13.79^{+2.61}_{-2.53}$\\
		      ABELL0115a & 0.1971 & 00 55 59.76 & $+$26 22 40.8 & 65 & 73 & $9.93^{+3.49}_{-3.49}$ & $14.34^{+2.22}_{-2.22}$ & $3.86^{+1.64}_{-1.33}$ & $14.98^{+3.07}_{-2.81}$\\
		      ABELL0209 & 0.2060 & 01 31 53.00 & $-$13 36 34.0 & 99 & 80 & $13.04^{+1.46}_{-1.46}$ & $19.73^{+2.12}_{-2.12}$ & $8.84^{+1.36}_{-1.23}$ & $26.49^{+2.78}_{-2.72}$\\
		      RXJ0142.0+2131 & 0.2803 & 01 42 02.64 & $+$21 31 19.2 & 57 & 67 & $7.87^{+1.93}_{-1.93}$ & $14.24^{+1.98}_{-1.98}$ & $4.07^{+0.86}_{-0.76}$ & $15.05^{+2.30}_{-2.24}$\\
		      ABELL0267 & 0.2300 & 01 52 48.72 & $+$01 01 08.4 & 25 & 31 & $6.74^{+1.44}_{-1.44}$ & $12.40^{+4.82}_{-4.82}$ & $3.30^{+0.69}_{-0.61}$ & $12.40^{+4.92}_{-4.90}$\\
		      ABELL0291 & 0.1960 & 02 01 44.20 & $-$01 12 03.0 & 42 & 61 & $7.55^{+1.56}_{-1.56}$ & $10.18^{+1.44}_{-1.44}$ & $4.11^{+1.00}_{-0.89}$ & $10.23^{+1.68}_{-1.63}$\\
		      ABELL0383 & 0.1883 & 02 48 02.00 & $-$03 32 15.0 & 56 & 87 & $7.59^{+1.61}_{-1.61}$ & $9.10^{+1.79}_{-1.79}$ & $3.39^{+0.73}_{-0.61}$ & $9.75^{+1.96}_{-1.92}$\\
		      ABELL0586 & 0.1710 & 07 32 22.32 & $+$31 38 02.4 & 76 & 71 & $10.78^{+3.46}_{-3.46}$ & $20.92^{+4.27}_{-4.27}$ & $6.77^{+2.00}_{-1.63}$ & $25.83^{+5.37}_{-5.16}$\\
		      ABELL0611 & 0.2880 & 08 00 55.92 & $+$36 03 39.6 & 64 & 72 & $10.22^{+1.94}_{-1.94}$ & $16.95^{+3.70}_{-3.70}$ & $5.19^{+1.00}_{-0.91}$ & $19.00^{+3.94}_{-3.91}$\\
		      ABELL0697 & 0.2820 & 08 42 57.84 & $+$36 21 54.0 & 77 & 83 & $11.91^{+1.62}_{-1.62}$ & $16.17^{+3.44}_{-3.44}$ & $8.39^{+1.27}_{-1.17}$ & $22.09^{+4.01}_{-3.99}$\\
		      ABELL1835 & 0.2528 & 14 01 02.40 & $+$02 52 55.2 & 127 & 91 & $15.70^{+2.94}_{-2.94}$ & $22.89^{+2.96}_{-2.96}$ & $9.69^{+1.71}_{-1.53}$ & $28.04^{+3.46}_{-3.37}$\\
		      ZwCl1454.8+2233 & 0.2578 & 14 57 14.40 & $+$22 20 38.4 & 40 & 78 & $7.07^{+2.89}_{-2.89}$ & $8.68^{+2.33}_{-2.33}$ & $2.61^{+0.99}_{-0.81}$ & $8.25^{+2.54}_{-2.47}$\\
		      ABELL2219 & 0.2281 & 16 40 22.56 & $+$46 42 21.6 & 113 & 80 & $11.10^{+2.26}_{-2.26}$ & $21.42^{+2.79}_{-2.79}$ & $8.10^{+1.50}_{-1.36}$ & $26.69^{+3.43}_{-3.36}$\\
		      RXJ1720.1+2638 & 0.1640 & 17 20 08.88 & $+$26 38 06.0 & 70 & 98 & $6.17^{+2.02}_{-2.02}$ & $10.44^{+2.05}_{-2.05}$ & $3.77^{+1.11}_{-0.94}$ & $10.79^{+2.32}_{-2.24}$\\
		      RXJ2129.6+0005 & 0.2350 & 21 29 37.92 & $+$00 05 38.4 & 40 & 70 & $8.37^{+1.83}_{-1.83}$ & $9.21^{+2.57}_{-2.57}$ & $4.69^{+1.10}_{-0.99}$ & $10.73^{+2.77}_{-2.74}$\\
		      ABELL2390 & 0.2329 & 21 53 36.72 & $+$17 41 31.2 & 122 & 85 & $13.75^{+1.99}_{-1.99}$ & $20.05^{+2.02}_{-2.02}$ & $7.10^{+1.29}_{-1.17}$ & $21.66^{+2.45}_{-2.39}$\\
		      ABELL2485 & 0.2472 & 22 48 31.13 & $-$16 06 25.6 & 51 & 85 & $7.74^{+2.39}_{-2.39}$ & $10.09^{+2.48}_{-2.48}$ & $3.29^{+0.90}_{-0.80}$ & $9.81^{+2.63}_{-2.60}$\\
		      \hline
		\end{tabular}
	\end{center}
{\footnotesize 
$\rm N_{gal}$: Number of spectroscopically confirmed member galaxies with $K\le K^{\ast}(\rm z)+1.5$. Completeness: Percentage of galaxies with $K\le K^{\ast}(\rm z)+1.5$ and within the $J{-}K$ colour cut that have spectroscopic data.}
\end{table*}

\subsection{Gravitational Weak-Lensing Masses}\label{sec:Mwl}

We use both model independent projected and model dependent deprojected weak-lensing masses from \citet{2010PASJ...62..811O} (Table~\ref{tab:sample}), in which Subaru/Suprime-Cam\footnote{Based in part on data collected at Subaru Telescope and obtained from the SMOKA, which is operated by the Astronomy Data Center, National Astronomical Observatory of Japan.} imaging was used to map the distribution of matter in each cluster. Details of the weak-lensing analysis can be found in \citeauthor{2010PASJ...62..811O} and are summarised here. Using deep $V$- and $i'$-band data, background galaxies were selected as those redder or bluer than the cluster red sequence \citep[following][]{2008ApJ...684..177U,2009ApJ...694.1643U}, and their redshifts estimated statistically by matching their colours and magnitudes to the COSMOS photometric redshift catalogue \citep{2009ApJ...690.1236I}. The KSB method \citep{1995ApJ...449..460K} was used to measure a shear estimate for each galaxy, by considering the PSF and residual mean ellipticity of point sources.

The model independent mass is estimated using aperture mass densitometry, as the azimuthally averaged tangential shear is related to the projected mass density. The $\zeta_c$-statistic \citep{2000ApJ...539..540C} relates the tangential shear to the 2D mass enclosed within a circular aperture. The 3D spherical mass, $M_{\Delta}$, is defined as the mass within radius $r_{\Delta}$, the radius within which the average density is $\Delta \times \rho_{crit}$, where $\rho_{crit}=3H(z)^2/8\pi G$, the critical density of the Universe. The values for $M_{\Delta}$ are estimated by fitting to the measured shear profile an NFW model parameterised by $M_{\Delta}$ and $c_{\Delta}$ (the concentration parameter), where $\rho(r) \propto(c_{\Delta}r/r_{\Delta})^{-1}(1+c_{\Delta}r/r_{\Delta})^{-2}$ \citep{1997ApJ...490..493N}.

We also consider the 3D spherical mass within a fixed radius, and the projected mass within the $r_{\Delta}$ values determined by the 3D analysis. We work with an overdensity $\Delta=500$ as $r_{500}$ is typically the limiting radius to which all mass measurement methods can probe, enabling comparisons, and a fixed radius of 1Mpc because $r_{500}\simeq1\rm Mpc$ for our sample.

Recent results \citep{2013A&A...550A.129P,2013ApJ...769L..35O,2014MNRAS.439...48A} suggest that \citeauthor{2010PASJ...62..811O}'s (\citeyear{2010PASJ...62..811O}) $M_{500}$ values may be underestimated by up to 20\%, with no obvious trend with mass. We therefore concentrate on the slope and scatter of the mass-luminosity relation. We will consider the absolute normalisation of the mass-luminosity relation and explore possible subtle systematics in the scatter and slope of the relation in our future article on the scaling relations of the full ``High-$L_X$'' LoCuSS sample (Smith et al., in prep.).

\begin{figure}
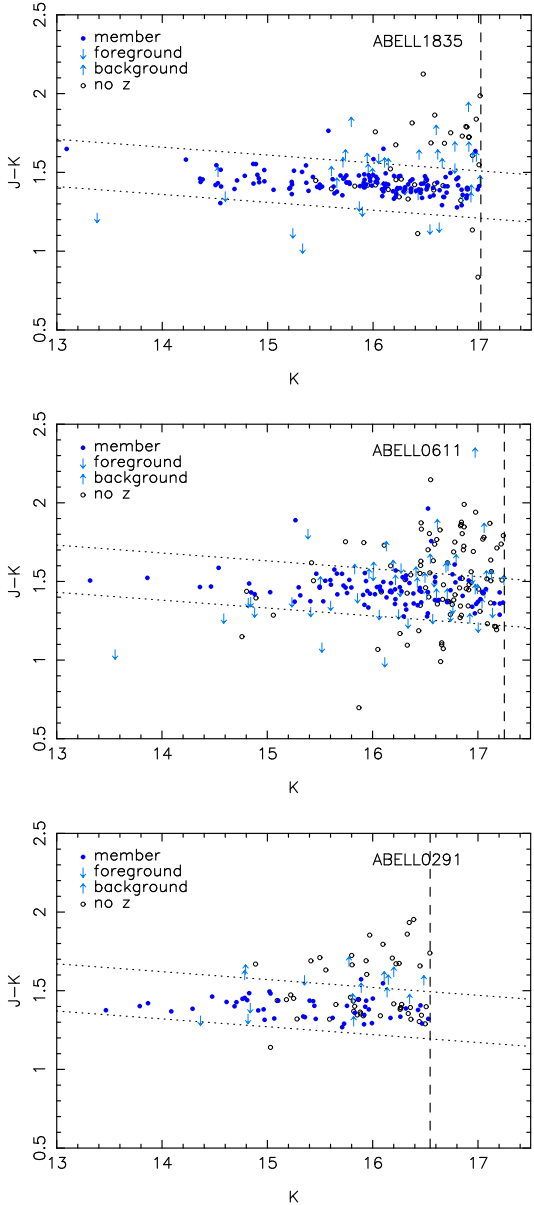

	\centerline{\psfig{file=figs/redseqhigh.ps,width=70mm,angle=-90}}
	\vspace{3mm}
	\centerline{\psfig{file=figs/redseqmid.ps,width=70mm,angle=-90}}
	\vspace{3mm}
	\centerline{\psfig{file=figs/redseqlow.ps,width=70mm,angle=-90}}
\caption{Colour-magnitude plots for three example clusters: the cluster with the highest number of galaxies above the magnitude cut $K^{\ast}(\rm z)+1.5$ - ABELL1835, the middle - ABELL0611, and the lowest - ABELL0291. The well defined ridge line of confirmed cluster members (dark blue filled points) can be clearly seen. Up and down arrows show background and foreground galaxies respectively, and hollow points show the galaxies with no spectroscopic data. The dotted lines show the width of the colour cut used for the colour selected $L_{K}$ measurements, and the vertical dashed lines mark $K^{\ast}(\rm z)+1.5$ for the respective cluster redshifts.}
\label{fig:redseq}
\end{figure}

\subsection{Observations}\label{sec:obs}

We have observed fifteen clusters from our sample with WFCAM on UKIRT\footnote{UKIRT is operated by the Joint Astronomy Centre on behalf of the Science and Technology Facilities Council of the United Kingdom.}, and the remaining two clusters with NEWFIRM on the Mayall 4-m telescope at Kitt Peak National Observatory\footnote{Kitt Peak National Observatory, National Optical Astronomy Observatory, which is operated by the Association of Universities for Research in Astronomy (AURA) under cooperative agreement with the National Science Foundation.}. Details of these observations can be found in \citet{2009ApJ...704..126H} and are summarised here. The WFCAM data cover $52'\times52'$ fields of view, while the NEWFIRM data consist of dithered and stacked images covering $27'\times27'$ fields of view, both to depths of $K\simeq19$, $J\simeq21$ with ${\rm FWHM}\simeq1''$. Total $K$-band Kron magnitudes were determined for each source, while $(J-K)$ colours were derived within fixed circular apertures of diameter $2''$.

Galaxy colours can be difficult to interpret, particularly in the optical, because they are affected by redshift, metallicity, star-formation rate and dust extinction. However, near-infrared wavelengths are relatively insensitive to the latter two, while $(J-K)$ evolves monotonically with redshift out to $z\sim0.5$. This means there is no distinction between the red sequence and the blue cloud; galaxies of a particular redshift lie along a single narrow relation in the $(J-K)/K$ colour-magnitude diagram (Figure~\ref{fig:redseq}), allowing us to simply select galaxies within a colour slice around this sequence in order to select all galaxies (passive and star-forming) within a redshift range centred on the cluster. This is in contrast to optical colour-magnitude diagrams which show a prominent blue cloud \citep[e.g.][]{2004ApJ...600..681B}, and a larger range of deviations from the red sequence within the cluster member population.

In addition to near-infrared data, we have spectroscopic data from MMT/Hectospec\footnote{Observations reported here were obtained at the MMT Observatory, a joint facility of the University of Arizona and the Smithsonian Institution.}, observed as part of the Arizona Cluster Redshift Survey (ACReS\footnote{\url{http://herschel.as.arizona.edu/acres/acres.html}}; M.\ J.\ Pereira et al.\ in preparation). The observation details can be found in \citet{2013ApJ...775..126H} and are summarised here. Hectospec is a 300-fiber multi-object spectrograph with a field of view of $1^{\circ}$ diameter on the 6.5m MMT telescope. The 270 line grating was used, providing a wide wavelength range (3650--9200{\AA}) at 6.2{\AA} resolution. Redshifts were determined by comparison of the reduced spectra with stellar, galaxy and quasar template spectra. Galaxies that fall within a colour slice around the ridge line of cluster members in the $(J-K)/K$ colour-magnitude diagram (Figure~\ref{fig:redseq}) were targeted by ACReS.

\subsection{Near-Infrared Luminosity}\label{sec:klum}

As with the mass measurements we calculate luminosities within both 1Mpc and $r_{500}$.

To determine which galaxies are in a cluster we plot their redshifts against distance from the centre of the cluster, which shows a trumpet shaped caustic profile as expected for galaxies infalling and orbiting within a massive gravitational structure. All galaxies within this caustic are identified as cluster members, and we select all those within a circular aperture (of radius $r_{500}$, and 1Mpc) on the sky. To account for spectroscopic incompleteness we weight each galaxy by the inverse probability of it having been observed spectroscopically. We give an initial equal weight (1.0) to all those galaxies which could have been targeted for spectroscopy. For each galaxy lacking a redshift, its weight is transferred equally to its ten nearest neighbouring galaxies on the sky with known redshift that had the same priority level in the targeting strategy.

Due to the magnitude limit of the spectroscopic coverage we only consider galaxies with $K\le K^{\ast}(\rm z)+1.5$, for which the average spectroscopic completeness is $75\%$ within 1Mpc (Table~\ref{tab:sample}). We base our estimates of $K^{\ast}(\rm z)$ on \citet{2006ApJ...650L..99L}. To convert from apparent $K$-band magnitude to rest frame luminosity, we use a $k$-correction consistent with \citet{2001MNRAS.326..745M}, and the absolute $K$-band magnitude of the sun, $M_{K,\odot}=3.39$ \citep{1966ARA&A...4..193J}. To account for the contribution of faint galaxies with $K > K^{\ast}(\rm z)+1.5$ we multiply the cluster luminosities by a factor of 1.286, calculated by assuming that the faint end of the cluster galaxy luminosity function has a slope of $\alpha=-1.0$ \citep[e.g.][]{2011MNRAS.412..947B}. 

We also use a second method to calculate cluster luminosity, which differs only in how cluster membership is determined. Spectroscopic data will not necessarily be available for large samples in future surveys, and so instead we use the $(J-K)/K$ colour-magnitude plots. Probable cluster members are identified as those lying within $\pm0.15$mags of the ridge line of cluster members in the $(J-K)/K$ colour-magnitude plots (Figure~\ref{fig:redseq}), and the luminosity calculation continues as above. We carry out a statistical background correction using two control fields (The UKIDSS-DXS Lockman Hole and XMM-LSS fields \citep{2007MNRAS.379.1599L}). For each cluster we place 30 apertures of radius matching that 
used for the cluster luminosity measurements, and perform the same colour selection and luminosity calculation. The mean and standard deviation on the background calculated in this way are subtracted from our cluster luminosity measurements and propagated into the error respectively. The colour selection identifies all but 48 ($<3\%$) of the confirmed members of the entire sample.

The error on the luminosity for each cluster is calculated from several components added in quadrature. The first, bootstrap resampling with replacement, involves calculating the cluster luminosity for $10^{5}$ resamples of its members, and the standard deviation of these luminosities is the error contribution. Another component, which is only valid for $L(<r_{\Delta})$, comes from the uncertainty in the radius, which comes from the uncertainty in the mass and causes an error in the luminosity.

The average of the ratio of luminosities calculated using both methods, $\left\langle L_{\rm spect.} / L_{\rm colour} \right\rangle$, is $0.97 \pm 0.06$ within $r_{500}$ and $0.98 \pm 0.06$ within 1Mpc; the consistency with unity showing the consistency between the methods
on average.

\begin{figure*}
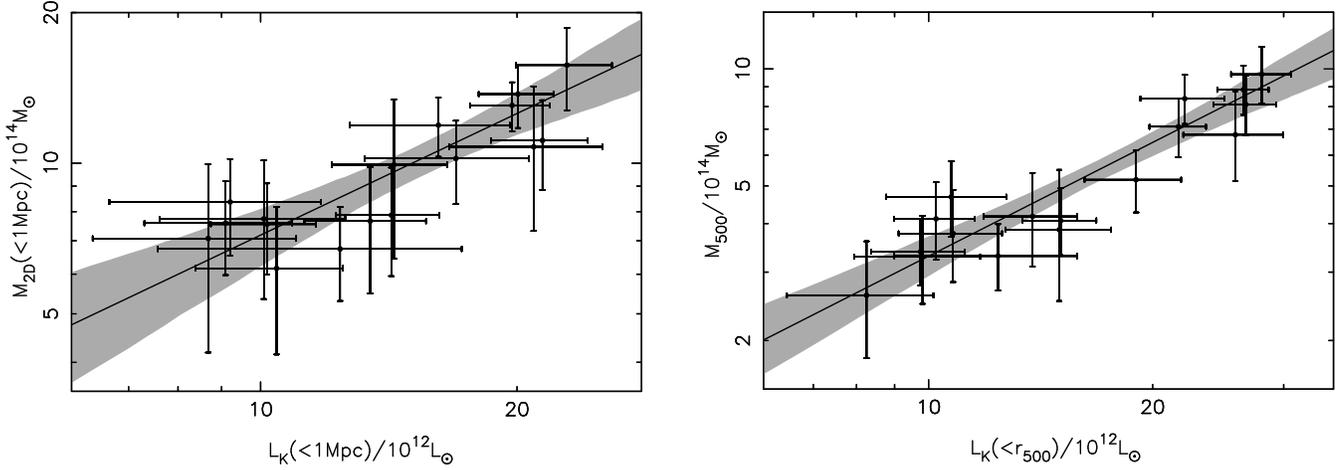

  \begin{minipage}{87mm}
    \psfig{file=figs/relationship1mpc2d.ps,width=85mm,angle=-90}
  \end{minipage}
  \hspace{5mm}
  \begin{minipage}{87mm}
    \vspace{1mm}
    \psfig{file=figs/relationship5003d.ps,width=85mm,angle=-90}
  \end{minipage}
  \caption{The scaling relation between weak-lensing mass and spectroscopically confirmed near-infrared luminosity. Left: the relation between projected mass and luminosity within a fixed metric aperture of 1Mpc. Right: the relation between the deprojected 3D mass and luminosity within $r_{500}$.}
  \label{fig:spectrel}
\end{figure*}

\begin{table*}
  \caption{Parameters for $M_{\rm WL} = a(L_{K})^b$ relations}\label{tab:relationvalues}
  \begin{center}
	\tabcolsep=0.8mm
    \begin{tabular}{ l c c c c c c }
      \hline
      \hline
      Member Selection & Radius & Normalisation & Slope & Intrinsic Scatter \\
      & & $(a)$ & $(b)$ & $(\sigma_{lnM_{\rm WL}|L_{K}}, \%)$ \\
      \hline
      \multicolumn{5}{c}{Model independent projected mass} \\
      Spectroscopic & 1Mpc & $1.06^{+0.98}_{-0.58}$ & $0.83^{+0.27}_{-0.24}$ & $10^{+8}_{-5}$ \\
      Spectroscopic & $r_{500}$ & $0.52^{+0.40}_{-0.25}$ & $0.96^{+0.22}_{-0.20}$ & $11^{+8}_{-6}$ \\
      Colour & 1Mpc & $0.84^{+1.02}_{-0.53}$ & $0.96^{+0.38}_{-0.30}$ & $11^{+8}_{-6}$ \\
      Colour & $r_{500}$ & $0.42^{+0.41}_{-0.22}$ & $1.10^{+0.27}_{-0.25}$ & $13^{+9}_{-7}$ \\
      \multicolumn{5}{c}{Model dependent deprojected mass} \\
      Spectroscopic & 1Mpc & $0.77^{+0.49}_{-0.35}$ & $0.80^{+0.23}_{-0.18}$ & $8^{+6}_{-4}$ \\
      Spectroscopic & $r_{500}$ & $0.35^{+0.22}_{-0.14}$ & $0.97^{+0.17}_{-0.17}$ & $10^{+7}_{-5}$ \\
      Spectroscopic & $r_{200}$ & $0.44^{+0.38}_{-0.19}$ & $0.92^{+0.20}_{-0.19}$ & $16^{+9}_{-8}$ \\
      Colour & 1Mpc & $0.62^{+0.51}_{-0.33}$ & $0.93^{+0.29}_{-0.24}$ & $10^{+7}_{-5}$ \\
      Colour & $r_{500}$ & $0.27^{+0.22}_{-0.14}$ & $1.12^{+0.26}_{-0.22}$ & $13^{+8}_{-6}$ \\
      Colour & $r_{200}$ & $0.25^{+0.29}_{-0.13}$ & $1.07^{+0.24}_{-0.24}$ & $16^{+10}_{-8}$ \\
      Spectroscopic & $L:\rm1Mpc$, $M:r_{500}$ & $0.19^{+0.21}_{-0.11}$ & $1.24^{+0.32}_{-0.27}$ & $13^{+9}_{-7}$ \\
      Colour & $L:\rm1Mpc$, $M:r_{500}$ & $0.13^{+0.22}_{-0.09}$ & $1.44^{+0.46}_{-0.45}$ & $15^{+11}_{-8}$ \\
      \hline
    \end{tabular}
  \end{center}{\footnotesize}
\end{table*}

\section{Results}\label{sec:results}

In this section we model the relation between mass and $K$-band luminosity. We measure the quantities within both 1Mpc and $r_{500}$, consider both 2D projected and 3D deprojected masses, and use luminosities based on both spectroscopic and colour member selection.

\subsection{Fitting Method}\label{sec:fitting}

To analyse the scaling relation between $M_{\rm WL}$ and $L_{K}$ we linearise the problem by taking the base-10 log of the respective measurements, and use a Bayesian approach to linear regression with a publicly available IDL code \citep{2007ApJ...665.1489K}. \citeauthor{2007ApJ...665.1489K} highlights the importance of correctly handling measurement errors when performing linear regression, and demonstrates that this model outperforms other estimators (OLS, BCES, FITEXY), especially when the measurement errors are large. The result of the routine is a line of best fit of the form:

\begin{equation}
  \frac{M_{\rm WL}}{10^{14}M_{\odot}}=a \left(\frac{L_{K}}{10^{12}L_{\odot}}\right)^{b},
\end{equation}
with normalisation $a$, slope $b$, and intrinsic scatter $\sigma_{\ln M_{\rm WL}|L_{K}}$.

\subsection{$M_{\rm WL}-L_{K}$ Relation}\label{sec:relations}

We first consider the relation between the 2D projected mass and luminosity within 1Mpc, because these quantities can be calculated directly from the data, with the fewest assumptions. Importantly, the use of a fixed metric aperture guarantees that the covariance between the mass and luminosity is zero.  We find a slope of $b=0.83^{+0.27}_{-0.24}$ and an intrinsic scatter of $\sigma=10^{+8}_{-5}\%$ (Figure~\ref{fig:spectrel}, Table~\ref{tab:relationvalues}).

The most common mass studied in the literature is the 3D overdensity mass $M_{\Delta}$.  We therefore also consider the scaling relation between deprojected mass and luminosity within $r_{500}$, both to enable comparisons with the literature, and because the halo mass function is typically expressed in terms of $M_{\Delta}$. We find the relation between 3D deprojected mass and luminosity within $r_{500}$ is parameterised by $b=0.97^{+0.17}_{-0.17}$ and $\sigma=10^{+7}_{-5}\%$ (Figure~\ref{fig:spectrel}, Table~\ref{tab:relationvalues}), again showing a promising low scatter.

We note that measuring the deprojected mass and luminosity within radii that scale with mass inevitably introduces covariance between the variables. In general, covariance may suppress the measured scatter in scaling relations.  However we draw attention to the consistency between the scatter measured for the relation between quantities inside a fixed metric aperture, and the result within $r_{500}$ above. This indicates that the impact of the covariance on the measured scatter is negligible.

From a cosmological perspective, the most meaningful mass measurement is the deprojected spherical mass $M_{500}$, however from an observational perspective, the simplest luminosity to measure is $L(<1\rm Mpc)$. We therefore fit a relation between these two values, finding $b=1.24^{+0.32}_{-0.27}$ and $\sigma = 13^{+9}_{-7}$ (Table~\ref{tab:relationvalues}). This relation is particularly important in demonstrating the potential of $L_{K}$ as a mass proxy for cluster cosmology, as measuring $L(<1\rm Mpc)$ does not require any prior radial information.

For completeness, we also measure the relations between deprojected mass and luminosity within 1Mpc, and projected mass and luminosity within $r_{500}$. We find that these relations also have low scatter, of $\sigma=8^{+6}_{-4}\%$ and $\sigma=11^{+8}_{-6}\%$ respectively, and that the slope of relations based on projected and deprojected mass are in close agreement (Table~\ref{tab:relationvalues}). Indeed, the slope of all of the spectroscopic relations is consistent with unity, and in agreement within the errors. 

However, we note that the central value of the slope of relations calculated within 1Mpc are consistently shallower than those calculated within $r_{500}$. Previous observational studies \citep[e.g.][]{1997ApJ...485L..13C,2004ApJ...610..745L,2005ApJ...633..122H} have shown that the number density profile of cluster galaxies is fit well by an NFW distribution, and \citet{2012MNRAS.423..104B} showed that the concentration parameter for the number density profile is a factor of two smaller than that of the dark matter density profile. This causes the stellar fraction to increase with cluster radius, following the same trend for all clusters relative to the overdensity radius. The fixed radius corresponds to a higher overdensity radius in larger clusters, and so results in a decreased stellar fraction, while the opposite is true for smaller clusters. This steepens the $L_{K}/M-M$ relation at 1Mpc compared to $r_{500}$, which leads to a shallower $M-L_{K}$ relation at 1Mpc compared to $r_{500}$.

Finally, we fit the scaling relation model to the same weak-lensing masses as discussed above, and near-infrared luminosities that are based on colour selection, as described in \S\ref{sec:klum}. We find that these colour selected scaling relations are fully consistent with the spectroscopically confirmed relations (Table~\ref{tab:relationvalues}).

\begin{table*}
  \caption{Comparison with Literature}\label{tab:literature}
  \begin{center}
	\tabcolsep=0.8mm
    \begin{tabular}{ l c c c c c c}
      \hline
      \hline
      Paper & Sample & Mass Measurement & Mass Range & Redshift Range & Slope & Intrinsic Scatter \\
      & Size & Technique & $(10^{14}M_{\odot})$ & & $(b)$& $(\sigma_{lnM|L_{K}} \%)$ \\
      \hline
      \multicolumn{7}{c}{$M_{500}$} \\
      \citet{2011MNRAS.412..947B} & 13 & X-Ray & $0.55 \le M_{500} \le 7.06$ & $0.05 \le z \le 0.095$ & $1.30^{+0.45}_{-0.46}$ & $64^{+22}_{-16}$ \\
      \citet{2004ApJ...610..745L} & 93 & $M-T_{X}$ Relation & $0.2 \le M_{500} \le 12.6$ & $0.016 \le z \le 0.09$ & $1.18^{+0.07}_{-0.07}$ & $25^{+6}_{-5}$ \\
      \citet{2003ApJ...591..749L} & 27 & $M-T_{X}$ Relation & $0.78 \le M_{500} \le 8.3$ & $0.016 \le z \le 0.09$ & $1.00^{+0.16}_{-0.16}$ & $28^{+9}_{-7}$ \\
      This work & 17 & Weak-Lensing & $2.6 \le M_{500} \le 9.7$ & $0.16 \le z \le 0.29$ & $0.99^{+0.21}_{-0.18}$ & $11^{+8}_{-6}$ \\
      \hline
      \multicolumn{7}{c}{$M_{200}$} \\
      \citet{2004AJ....128.2022R}$^b$ & 55 & Velocity Dispersion & $0.007 \le M_{200} \le 10.23$ & $z \le 0.04$ & $1.21^{+0.14}_{-0.14}$ & $57^{+10}_{-9}$ \\
      \citet{2004AJ....128.2022R}$^c$ & 61 & Velocity Dispersion & $0.007 \le M_{200} \le 15.49$ & $z \le 0.05$ & $1.21^{+0.09}_{-0.09}$ & $50^{+9}_{-8}$ \\
      \citet{2004AJ....128.1078R} & 9 & Caustics & $0.76 \le M_{200} \le 7.8$ & $z \le 0.05$ & $1.17^{+0.36}_{-0.30}$ & $29^{+21}_{-15}$ \\
      \citet{2004AJ....128.2022R}$^a$ & 36 & Velocity Dispersion & $0.039 \le M_{200} \le 10.23$ & $z \le 0.04$ & $1.12^{+0.25}_{-0.25}$ & $62^{+13}_{-12}$ \\
      \citet{2007ApJ...663..150M} & 14 & Velocity Dispersion & $3.5 \le M_{200} \le 33.3$ & $0.17 \le z \le 0.54$ & $1.08^{+0.29}_{-0.29}$ & $45^{+17}_{-12}$ \\
      \citet{2004ApJ...610..745L} & 93 & $M-T_{X}$ Relation & $0.3 \le M_{200} \le 18.9$ & $0.016 \le z \le 0.09$ & $1.07^{+0.06}_{-0.06}$ & $28^{+5}_{-4}$ \\
      This work & 17 & Weak-Lensing & $4.0 \le M_{200} \le 15.5$ & $0.16 \le z \le 0.29$ & $0.93^{+0.20}_{-0.19}$ & $16^{+10}_{-8}$ \\
      \citet{2011MNRAS.412..947B}$^d$ & 18 & Velocity Dispersion & $1.66 \le M_{200} \le 5.97$ & $0.05 \le z \le 0.096$ & $0.05^{+0.18}_{-0.17}$ & $24^{+12}_{-10}$ \\
      \hline
    \end{tabular}
  \end{center}
      {\footnotesize $^a$Core sample. $^b$Total sample. $^c$Extended sample, inc 5 \citet{2004AJ....128.1078R} groups/clusters. $^d$The shallow slope of this sample is likely a consequence of limiting the dynamic range in the dynamical mass, as noted by the authors.}
\end{table*}

\section{Discussion}\label{sec:discussion}

In \S\ref{sec:literature-m} we compare our results with other weak-lensing based mass-observable scaling relations, and in \S\ref{sec:literature-lk} we compare our results with previous measurements of the scaling relation between mass and near-infrared luminosity.

\subsection{Comparison with previous weak-lensing based scaling relation results}\label{sec:literature-m}

Our results, based on a small pilot study sample, show that the intrinsic scatter in the scaling relation between weak-lensing mass and near-infrared luminosity is $\simeq10\%$ on scales of 1Mpc, which corresponds to an overdensity of 500 with respect to the critical density of the Universe. This result is independent of whether the scaling relation is derived from measurements within a fixed metric aperture, or within a radius ($r_{500}$) that scales with mass, and independent of whether the luminosity is based on spectroscopically confirmed members or galaxies selected in the $(J-K)/K$ colour-magnitude plane. The scatter in weak-lensing mass to near-infrared luminosity scaling relation is therefore smaller than than that found in all previous weak-lensing-based studies of mass-observable scaling relations \citep{2010ApJ...721..875O,2012ApJ...754..119M,2013ApJ...767..116M}, with the exception of \citeauthor{Hoekstra2012}'s (\citeyear{Hoekstra2012}) relation between mass and the integrated Compton $Y_{SZ}$ parameter.

These results all point to observables that are closely related to a line-of-sight integral of a linear quantity through the cluster potential being low scatter proxies for the weak-lensing mass of clusters. Arguably the projected near-infrared luminosity of a cluster within a fixed metric aperture is the least expensive and least model dependent of the available observables because it is based on simply measuring flux from galaxies above a well-defined limit, and is feasible with wide-field survey data.

\subsection{Comparison with previous studies of $M-L_{K}$}\label{sec:literature-lk}

\begin{figure}
 \centerline{\psfig{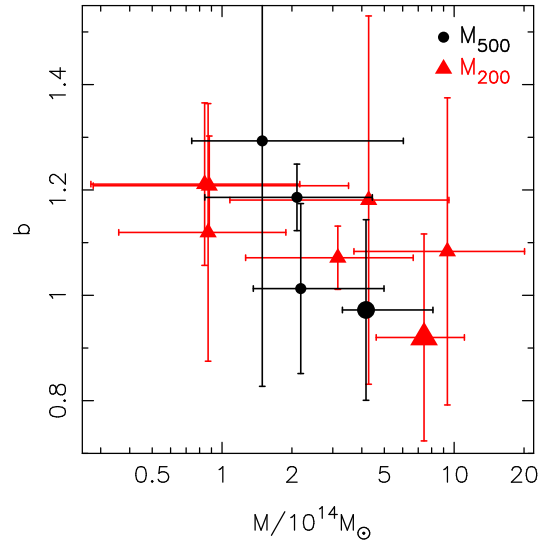}}
  \caption{The slopes of the $M = aL_K^b$ relation fit to each literature sample using the \citet{2007ApJ...665.1489K} method (Table~\ref{tab:literature}). The points show the slope against the average mass, the vertical error bars show the error on the slope and the horizontal error bars enclose $68\%$ of the mass range. The large points are the results from this work using spectroscopic member selection and 3D NFW masses. Note that the \citet{2011MNRAS.412..947B} $\Delta=500$ data is not visible as the slope is much shallower than the other results.}
  \label{fig:gradcomp}
\end{figure}

\begin{figure*}
    \begin{minipage}{86mm}
      \psfig{file=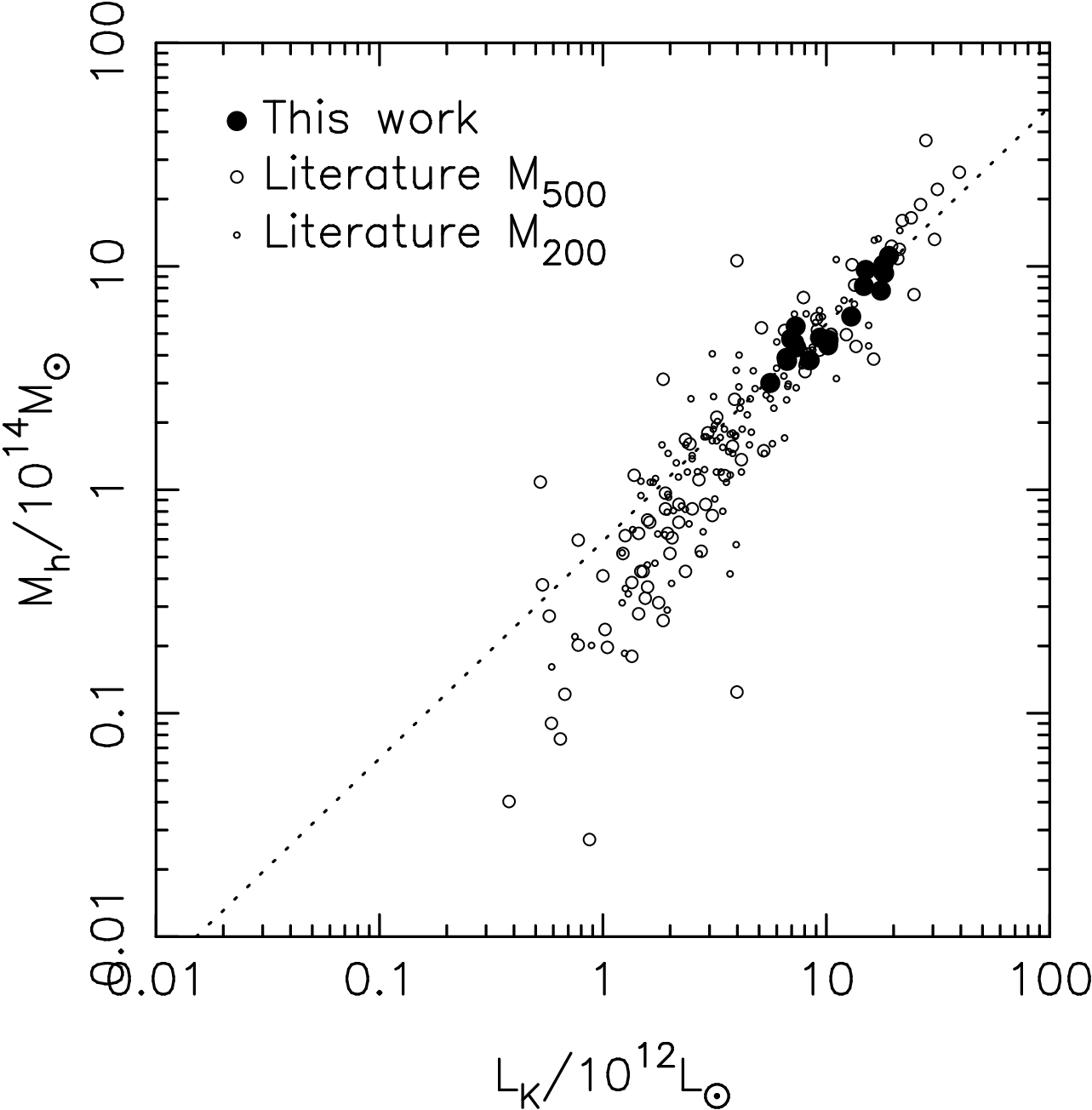,width=86mm,angle=-90}
	\caption{Comparison with all the available data from the literature, normalised such that the individual scaling relations overlap with our $M_{\rm WL,500}$ relation (dotted line) at our mean $M_{500}$ value. For the samples analysed in two papers and/or at two overdensities, we plot the most recent and/or highest overdensity values.}
  \label{fig:plotall}
    \end{minipage}
    \hspace{1.5mm}
    \begin{minipage}{86mm}
    \vspace{4mm}
      \psfig{file=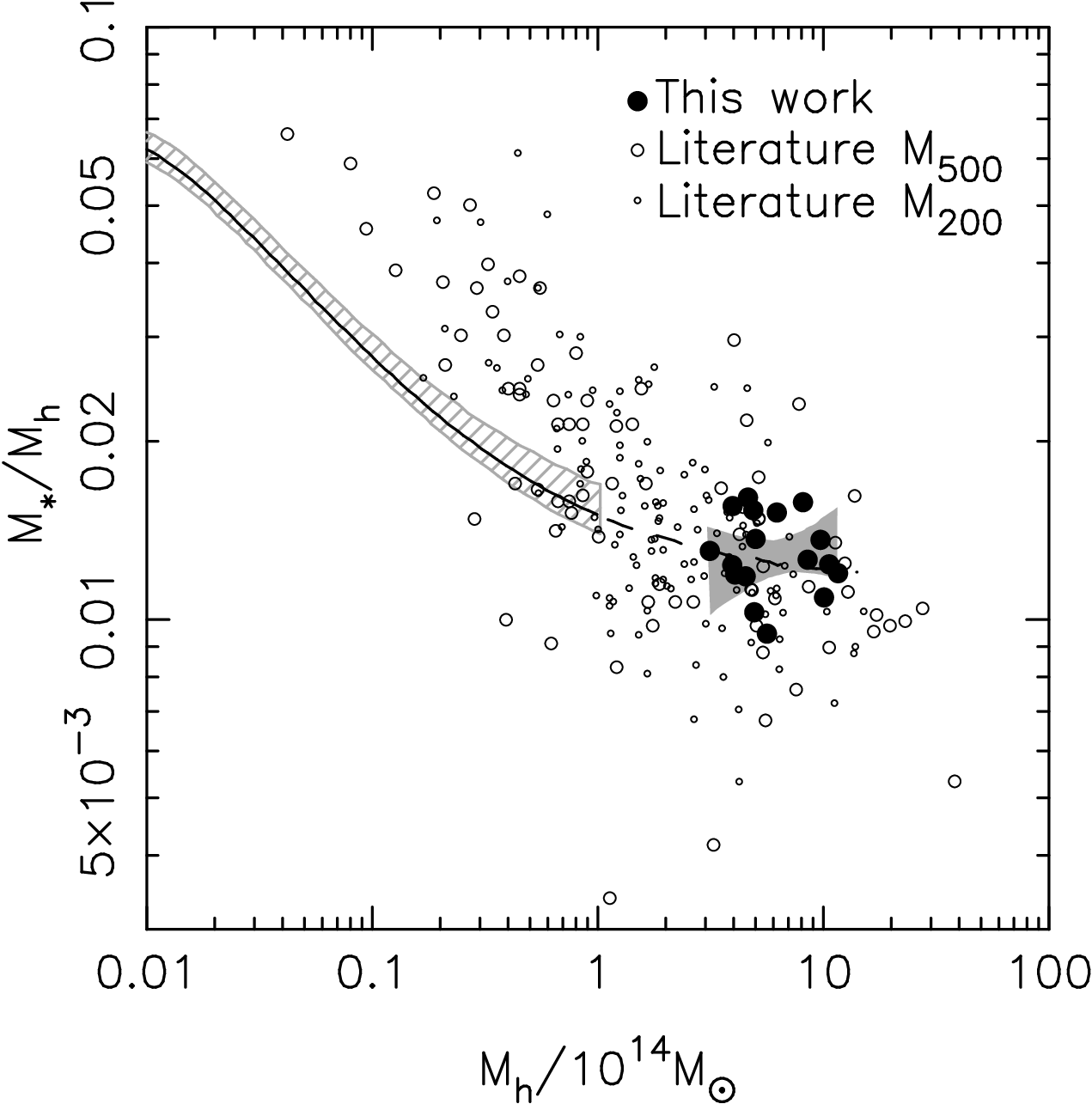,width=86mm,angle=-90}
  \caption{The data from Figure~\ref{fig:plotall}, where $L_{K}$ has been converted to $M_{\ast}$ using a mass to light ratio of 0.73 \citep{2001MNRAS.326..255C}. Also shown is the trend and error envelope from \citeauthor{2012ApJ...746...95L}'s (\citeyear{2012ApJ...746...95L}) halo occupation distribution model within $r_{500}$; where the dashed line shows the extrapolation beyond the data. The solid shaded region shows the error envelope from our $M_{500}-L_{K}$ relation.}
  \label{fig:stellar_mass}
    \end{minipage}
\end{figure*}

The intrinsic scatter in our $M_{500}-L_{K}$ relation is much lower than the scatter of $\sigma_{lnL_{K}|M} = 28\%$ found by \citet{2003ApJ...591..749L}. These authors estimated $M_{500}$ from the relationship between hydrostatic mass and X-ray temperature.  To compare our work more directly with \citeauthor{2003ApJ...591..749L} we repeat our fit of the $M_{500}-L_K$ relation using hydrostatic masses \citep{2014arXiv1406.6831M} in place of our weak-lensing masses.  We measure an intrinsic scatter of $\sigma_{lnM|L_{K}} = 25^{+11}_{-10}\%$, which is consistent with \citeauthor{2003ApJ...591..749L}'s result, and supports the interpretation of the weak-lensing based scaling relation results discussed in \S\ref{sec:literature-m}.

We now concentrate on comparing our $M_{WL}-L_K$ results with those in the literature, and make two corrections to ensure that our data are comparable. Firstly, we multiply our $M_{500}$ values by 1.20, to account for the 20\% bias in the mass measurements as discussed in \S\ref{sec:Mwl}. Secondly, we deproject our luminosities. Counting all the cluster members within $r_{\Delta}$ on the sky gives a cylinderical volume projected along the line of sight within which we calculate the luminosity, which requires deprojection to correct to a spherical volume. We therefore multiply our $L_{500}$ values by 0.68, the average ratio of the 3D to 2D $M_{500}$ measurements \citep{2010PASJ...62..811O}. We use a constant based on the NFW profile for this deprojection, as do \citet{2007ApJ...663..150M} (0.791), \citet{2004AJ....128.2022R} (0.80) and \citet{2009ApJ...703..982G} (0.86). After these corrections our error weighted mean mass-to-light ratio is $55.9 \pm 1.8 M_{\odot}/L_{\odot}$, which is consistent with other results in the literature \citep[e.g.][]{2001ApJ...561L..41R,2003ApJ...585..161K,2003ApJ...591..749L,2004AJ....128.1078R,2007ApJ...663..150M}.

Results in the literature are generally expressed as $L_{K} = a\,M^b$, and in that form the slope of our deprojected mass relation within $r_{500}$ is $b = 1.00^{+0.21}_{-0.18}$.  The published results are generally shallower than this (equivalent to steeper in the form $M = aL_{K}^b$). To ensure that this is not caused by a difference in fitting method, we refit each sample in the published literature with the \citet{2007ApJ...665.1489K} method in the same manner as our results in \S\ref{sec:results}, in the form $M = aL_{K}^b$ (Table~\ref{tab:literature}). We find that the flatter slope of our $M-L_K$ relation is not an artefact of fitting method. However we notice that in general the dynamic range of mass explored by other authors is wider than our own, and extends to lower masses. This suggests that the slope of the $M-L_K$ relation may be a function of halo mass (Figure~\ref{fig:gradcomp}).

To further illustrate this point, we plot all the available data from the literature after re-normalising it with respect to our own, as we are focussing on the slope of the relation. For each sample we calculate the normalisation required to make the relevant best fit scaling relation intersect our relation at the mean mass of our sample, and apply that normalisation adjustment to every cluster in that sample (Figure~\ref{fig:plotall}).

We caution that the general shallowing of the $M-L_K$ relation may be an artefact of selection and/or measurement biases at low mass. Nevertheless, taking the gradual shallowing at face value corresponds to a smaller stellar fraction for larger clusters. To explore this further we use a simple method to calculate $f_* \equiv M_*/M_h$ for our sample and compilation from the literature using a stellar mass-to-light ratio of 0.73 \citep{2001MNRAS.326..255C}. For comparison we show the results from \citeauthor{2012ApJ...746...95L}'s (\citeyear{2012ApJ...746...95L}) halo occupation distribution model within $r_{500}$, noting that the same trend is found using abundance matching techniques \citep[e.g.][]{2010MNRAS.404.1111G,2013MNRAS.428.3121M,2013ApJ...770...57B,2014arXiv1401.7329K}. The decreasing stellar fraction seen in Figure~\ref{fig:stellar_mass} suggests a quenching of star formation in larger systems, which is consistent with results of other observational studies \citep[e.g.][]{2007ApJ...666..147G,2011ApJ...743...13L}. We also note that our results on the slope of the mass-luminosity relation of clusters -- i.e.\ a linear relation between weak-lensing mass and $K$-band luminosity -- suggest that for the most massive halos the relationship between stellar mass fraction and halo mass may be flatter than implied by an extrapolation of by \citeauthor{2012ApJ...746...95L}'s relation. This emphasises the importance of direct calibration of this relation, as highlighted recently by \citet{2014arXiv1401.7329K}.

\section{Summary}\label{sec:summary}

In this pilot study we have shown that $K$-band luminosity is a promising low scatter proxy for weak-lensing mass, with an intrinsic scatter of $\sim 10\%$.

A useful mass proxy must be easy to measure, and so we have considered the values closest to the data plane - projected values within 1Mpc - and found a scatter of only $\sigma_{lnM_{\rm WL}|L_{K}}=10^{+8}_{-5}\%$, demonstrating the practical potential of the relation. We have also shown that having spectroscopic information is not required, as the scatter does not increase when determining cluster membership using the $(J-K)/K$ colour-magnitude diagram. It will not be practical to have such spectroscopic coverage for future surveys, and so this is an important result.

The halo mass function is typically expressed in terms of $M_{\Delta}$, so it is also of interest to study the scaling relation between mass and luminosity estimated within the three-dimensional over-density radius. We therefore considered the relation between deprojected $M_{500}$ and $L_{K}(<r_{500})$ and found an intrinsic scatter of only $\sigma_{lnM_{\rm WL}|L_{K}}=10^{+7}_{-6}\%$. We also note that the invariance of the scatter between the relation measured within a fixed metric aperture and that measured within $r_{500}$ indicates that the impact of covariance between mass and luminosity via the use of $r_{500}$ in the latter relation has negligible effect on the measured scatter.

The above relation was motivated by the most useful mass quantity for cosmology, while the first relation we considered was motivated by the most practical luminosity to measure. We combined the advantages of both these relations by considering the relation between $M_{500}$ and $L(<1\rm Mpc)$. The resulting low scatter of only $\sigma_{lnM_{\rm WL}|L_{K}}=13^{+9}_{-7}\%$ demonstrates the potential of $L_{K}$ as a mass proxy for cluster cosmology. This highlights the importance of calibrating the relation as a function of both redshift and mass.

The studies in the literature against which we compared our results used a range of mass measurements and find consistently higher scatter than our $\simeq10\%$, suggesting that the low intrinsic scatter in the $M_{\rm WL}-L_{K}$ relation is related to both quantities suffering similar projection effects. When compared to the literature there appears to be a mass dependence in the slope of the relation; the slope of the $M-L_{K}$ relation appears to be a decreasing function of mass. This is equivalent to a stellar fraction $M_*/M_h$ that decreases with increasing mass, suggesting a quenching of star formation in larger systems.

Encouraged by the positive result of this pilot study, in future work we will investigate this relation for a statistically complete sample of 50 clusters for which we now have near-infrared data -- the LoCuSS ``High-$L_X$'' sample. With this larger sample, and improved weak-lensing masses, we will be able to reduce statistical errors and subtle biases in our results and also investigate the effect of cluster morphology on the relation.  We expect that our results will be helpful for upcoming large-scale optical/infrared surveys that will study galaxy clusters, with cosmological goals, including HSC, DES, Euclid, and LSST.

\section*{Acknowledgments}

SLM acknowledges support from an STFC Postgraduate Studentship. SLM and GPS acknowledge support from the Royal Society. GPS and CPH acknowledge support from STFC. CPH was funded by CONICYT Anillo project ACT-1122. We thank Trevor Ponman, Felicia Ziparo, Keelia Scott and Maggie Lieu for helpful discussions and suggestions.

\bibliographystyle{mn2e}
\bibliography{Mulroybib}

\label{lastpage}
\end{document}